\documentclass[11pt]{article}
\pdfoutput=1
\usepackage{graphicx}
\usepackage{color}
\usepackage{amsmath}
\usepackage{amssymb}
\usepackage{slashed}
\usepackage{epsfig}
\usepackage{textcomp}
\usepackage{hyperref}
\usepackage{geometry}
\newcommand{\eps}{\epsilon}
\newcommand{\gpZ}{g g \to  \gamma Z g}
\newcommand{\gVV}{g g \to  V V^\prime g}
\newcommand{\VV}{g g \to  V V^\prime }

\newcommand{\gpp}{g g  \to \gamma \gamma g}
\newcommand{\gww}{g g  \to W^+ W^- g}

\begin{document}

\title{\bf Di-Vector Boson + Jet Production via Gluon Fusion at Hadron Colliders}
\author{{Pankaj Agrawal \footnote{email:agrawal@iopb.res.in}}~~and
       {Ambresh Shivaji \footnote{email:ambresh@iopb.res.in}} \\
{Institute of Physics, Sainik School Post, Bhubaneswar 751 005, India} }

\date{}

\maketitle{}


\begin{abstract}
 { We study the production of two electro-weak vector bosons in association
 with a jet via gluon fusion. In particular we consider the production of
 $\gamma Zg$, $ZZg$ and $W^{+}W^{-}g$ at hadron colliders and compute their 
 cross-sections.  Such processes have already produced large number
 of events at the Large Hadron Collider (LHC) by now. These processes can be background to
 the Higgs boson production and a number of beyond the standard model
 scenarios.  Therefore it is important to know the values of their
 contribution. The calculation is based on conventional Feynman diagram
  approach. In particular we find that the process $g g \to Z Z g$ can
  make a significant contribution to the process $p p \to Z Z + {\rm jet}$.}
\end{abstract}


\section{Introduction}

   The search for new physics at the LHC is in progress and the collider is delivering
   data presently at 8 TeV centre-of-mass energy. The discovery of a fundamental
   scalar particle (most probably a Higgs boson) of mass around 125 GeV has received a 
   lot of world-wide attention \cite{cern}. We expect more good news from the experiments at 
   the LHC before the collider goes for a two years long pause. So far, the standard 
   model (SM) of particle physics
  seems to be in excellent agreement with the collected data (more than $10\;{\rm fb}^{-1})$. 
  There has been searches for the hints of physics beyond the SM such as supersymmetry, large extra 
  dimensions, etc. But, as of now, there is no clear evidence \cite{talks}. The process of 
  identifying the discovered fundamental scalar particle as \emph{the Higgs boson} is also continuing. 
  With the lack of signals for beyond the SM scenarios, there is a need to look for the SM 
  processes that were not accessible earlier at the Tevatron. Most of such processes have several 
  particles in the final state, and/or occur at the higher order. Such processes can also
  contribute to the background to the new physics signals. One such class of processes is multi 
  vector boson production in association with one or more jets.

\begin{figure}[h!]
\begin{center}
 \includegraphics[width=0.5\textwidth]{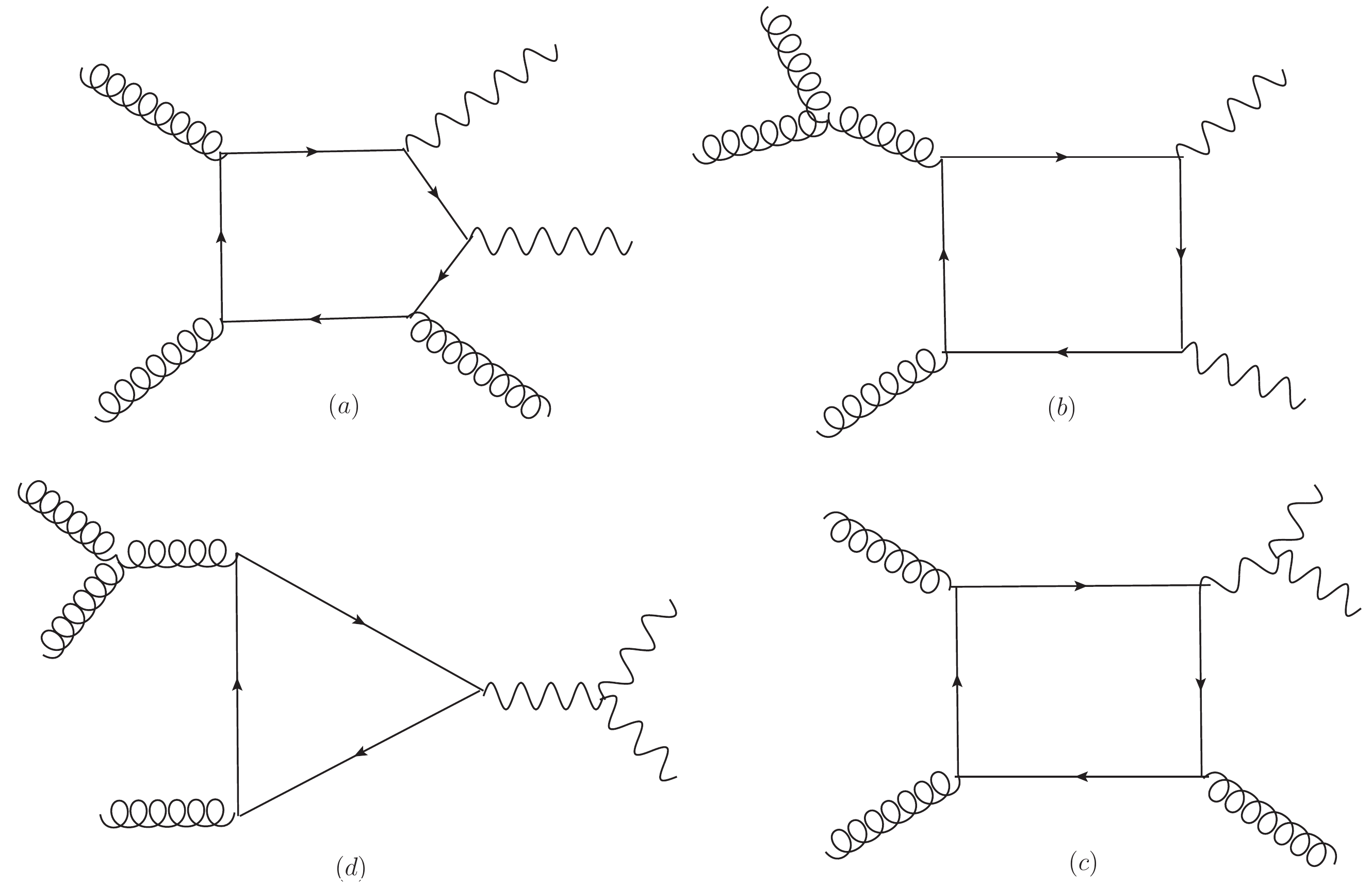}
\caption{The prototype diagrams for the processes $gg \to V V^\prime g$. 
The wavy lines represent the appropriate combination of
the $\gamma, W, \rm{or}\; Z$ boson. The last two classes $(c)$ and $(d)$ are relevant to $WWg$
production only. We do not consider diagrams involving Higgs boson for the $ZZg$ and $WWg$ cases.}
\label{fig:VVg}
\end{center}
\end{figure}

    At the LHC centre-of-mass energy, the collider has another useful feature.
   In the proton-proton collisions, the gluon luminosity can be quite significant.
   It can even dominate over the quark luminosity in certain kinematic domains.
   Therefore, at the LHC, loop mediated gluon fusion processes can be important.
   Di-vector boson production via gluon fusion have been studied by many authors 
   \cite{Ametller:1985di,Dicus:1987dj,Dicus:1987fk,vanderBij:1988fb,Glover:1988rg,Glover:1988fe,Matsuura:1991pj,
   Zecher:1994kb,Binoth:2005ua,Binoth:2006mf,Binoth:2008pr,Campbell:2011bn,Campbell:2011cu}.
   We consider another class of processes $gg \to V V^\prime g$, where $V$ and $V^\prime$
   can be any allowed combination of electro-weak vector bosons. These processes can be
   a background to Higgs boson production as well as new physics scenarios such as 
   technicolor. At the leading order, these processes
   receive contribution from quark loop diagrams. The prototype diagrams are displayed
   in Fig (\ref{fig:VVg}). The computation for the process $\gpp$ has already been performed 
   \cite{deFlorian:1999tp,Agrawal:1998ch}. Preliminary results for $\gpZ$ were presented in \cite{Agrawal:2012sq}.
    Recently, {\it Melia et al.} have presented calculation for $\gww$ \cite{Melia:2012zg}.
   In this brief report we present our results
   for the production of $\gamma Zg$, $ZZg$ and $W^{+}W^{-}g$ at hadron colliders. Detailed
   analysis of these processes will be reported in future publications.


\section{The Process: $gVV^\prime$ }

   The process $\gVV$ contribute to the cross-section of the process 
   $pp \to V V^\prime + \rm{jet}$ at NNLO in $\alpha_s$. As discussed,
   although they occur at the one-loop level, their cross-sections can be significant
   due to large gluon luminosity at the LHC. The processes $\gamma Z g$ and $ZZg$ receive contribution 
   from two main classes of quark-loop diagrams -- pentagon and box type, as shown 
   in $(a)\;\rm{and}\;(b)$ of Fig (\ref{fig:VVg}). 
   Box class of diagrams are due to the triple gluon vertices; they
   can be further divided into three sub-classes. This sub-classification has it's 
   own physical importance. These are separately gauge invariant with respect to the electro-weak
   vector bosons. Other diagrams can be obtained by suitable permutation of external
   legs. For each quark flavor, there are 24 pentagon-type and 18 box-type diagrams. 
   Due to Furry's theorem, only half of the 42 diagrams are independent. Also pentagon diagrams give 
   both vector as well as axial-vector contributions while box diagrams give only vector contribution.
   We work with five massless quark flavors and expect decoupling of the top quark.

   In the case of $WWg$ process, instead of single quark flavor, two quark flavors of a single generation contribute
   to the above discussed pentagon and box diagrams.
   To keep matter simple, for this process, we work with first two generations of massless quarks.
   It is expected that the contribution from the third generation would not be significant in  
   low $P_T$ region \cite{Campbell:2011cu}. There are additional box and triangle class of 
   diagrams due to $\gamma/Z\;W^{+} W^{-}$ vertex for each quark flavor. These are shown in $(c)$ and $(d)$
    of Fig. (\ref{fig:VVg}). Due to Furry's theorem, the triangle
   diagrams with $\gamma W^{+} W^{-}$ coupling don't contribute and only axial part of the triangle diagrams with
   $ZW^{+} W^{-}$ coupling contribute to the amplitude. Since the axial coupling of $Z-$ boson to quark
   is proportional to $T^{(q)}_3$ value, the axial contributions from additional triangle and box diagrams,
   when summed over massless quark generation, vanish. The vector contribution of additional 
   box-type diagrams is separately gauge invariant. Because of its color structure, it interferes with
   the axial part of the pentagon amplitude. We have explicitly checked that its contribution towards
   total cross-section is very small; therefore we have dropped this contribution.
   Thus, effectively we are left with $ZZg$  like contributions for $WWg$. The 
   Higgs boson interference effects for the cases of $ZZg$ and $WWg$ are ignored in the present calculation. 
    Our one loop processes, being the leading order processes, are expected to be finite, i.e. free from
    UV and IR divergences.

  The amplitude for our process has the following general structure 

\begin{eqnarray}
{\cal M}^{abc}(\gVV) & = & i \frac{f^{abc}}{2} {\cal M}_V(V V^\prime g) + \frac{d^{abc}}{2} {\cal M}_A(V V^\prime g), 
        \nonumber \\
{\cal M}_V(V V^\prime g) & = & -\; e^2 g_s^3 \; C_V(V V^\prime g)\;\left( {\cal P}_V - {\cal B}_V \right), \nonumber \\
{\cal M}_A(V V^\prime g) & = & -\; e^2 g_s^3 \; C_A(V V^\prime g)\;\left( {\cal P}_A \right).
\end{eqnarray}

${\cal M}_{V,A}$ are amplitudes for the vector and axial-vector parts of the full amplitude under consideration. Because 
of Bose symmetry ${\cal M}_V \rightarrow -{\cal M}_V$ under exchange of any two external gluons while ${\cal M}_A$ remains
same. ${\cal B}_V\;\rm{and}\; {\cal P}_{V,A}$ are box and pentagon contributions from a single flavor (single generation for $WWg$ case) of quarks.
The structure of the amplitude suggests that the vector and axial-vector contributions should be separately gauge invariant.
Moreover due to the color structure when we square the amplitude, the interference between the axial and vector 
contribution vanishes, $i.e.$
\begin{eqnarray}
|{\cal M}(\gVV)|^2 = \left( 6 |{\cal M}_V|^2 + \frac{10}{3} |{\cal M}_A|^2 \right).
\end{eqnarray}
     
Therefore the cross-section of the process is an incoherent sum of the vector and axial-vector contributions. 
The couplings $C_{V,A}$ for various cases are listed below. Contributions from all the relevant quark flavors
(described above) are included appropriately. 

\begin{eqnarray}
C_V(\gamma \gamma g) & = & \frac{11}{9}, \;\; C_A(\gamma \gamma g) = 0, \nonumber \\ 
C_V(\gamma Z g) & = & \frac{1}{sin\theta_w cos\theta_w} \left( \frac{7}{12} - \frac{11}{9} sin^2\theta_w \right),
\nonumber \\ 
C_A(\gamma Z g) & = & \frac{1}{sin\theta_w cos\theta_w} \left( -\frac{7}{12} \right),\nonumber  \\
C_V(Z Z g) & = & \frac{1}{sin^2\theta_w cos^2\theta_w} \left( \frac{5}{8} - \frac{7}{6} sin^2\theta_w +
                 \frac{11}{9} sin^4\theta_w \right), \nonumber \\ 
C_A(Z Z g) & = & \frac{1}{sin^2\theta_w cos^2\theta_w} \left( -\frac{5}{8} + \frac{7}{6} sin^2\theta_w \right), \nonumber \\
C_V(W W g) & = & \frac{1}{sin^2\theta_w} \left( \frac{1}{2} \right),\nonumber \\ 
C_A(W W g) & = & \frac{1}{sin^2\theta_w} \left( -\frac{1}{2} \right).
\end{eqnarray}

\section{Calculation and Numerical Checks}

 For each class of diagrams, we write down the prototype amplitudes using the SM Feynman rules \cite{peskin:2005}. 
 The amplitude of all other diagrams are generated by appropriately permuting the external momenta and 
 polarizations in our code. The quark loop traces without $\gamma_5$ are calculated in $n$-dimensions while
 those with $\gamma_5$ are calculated in $4$-dimensions using FORM \cite{Vermaseren:2000nd}. We do not need
 any $n$-dimensional prescription for $\gamma_5$ as both pentagon and box diagrams are free of anomaly. 
 The amplitude contains tensor loop integrals. In the case of pentagon-type diagrams, the most complicated integral
 is rank-5 tensor integral ($E^{\mu \nu \rho \sigma \delta}$); while for the box-type diagrams, rank-4 tensor 
 integral ($D^{\mu \nu \rho \sigma}$) is the most complicated one 

\begin{equation}
E^{\mu \nu \rho \sigma \delta} = \int \frac{d^n k}{(2 \pi)^n} \frac{k^{\mu} k^{\nu} k^{\rho} k^{\sigma} k^{\delta}}
 {N_0 N_1 N_2 N_3 N_4}\,,
\end{equation}

\begin{equation}
D^{\mu \nu \rho \sigma} = \int \frac{d^n k}{(2 \pi)^n} \frac{k^{\mu} k^{\nu} k^{\rho} k^{\sigma}}
 {N_0 N_1 N_2 N_3}\,.
\end{equation}

Here, $N_i = k_i^2 - m_q^2 + i\eps$ and $k_i$ is the momentum of the $i^{th}$ internal line in the
corresponding scalar integrals. $n=(4-2\eps)$ and $m_q$ is the mass of the quark in the loop.
Five point tensor and scalar integrals are written in terms of box tensor and scalar integrals using 
4-dimensional Schouten Identity. For example, a five point scalar integral, in 4-dimensions, can be 
expressed in terms of five box scalar integrals \cite{vanNeerven:1983vr}

\begin{equation}
 E_0(0,1,2,3,4) = \sum\limits_{i=0}^4 c_i D_0^{(i)},
\end{equation}

where $D_0^{(i)}$ is the box scalar integral obtained after removing the $i^{th}$ propagator
in $E_0$.  Any $O(\eps)$ correction to the above relation in $n=(4-2\eps)$-
dimensions does not contribute to our amplitude. This is because the contributions of the 
pentagon-type diagrams are UV finite and we have regulated mass singularities by giving
small mass to the quarks.
The box tensor integrals are reduced into the standard scalar integrals -- $A_0$, $B_0$, $C_0$ and $D_0$
 using FORTRAN routines that follows from the reduction scheme
developed by Oldenborgh and Vermaseren \cite{vanOldenborgh:1989wn}.
 Thus we require box scalar integrals with two massive external legs, at the most.
The scalar integrals with massive internal lines are calculated using OneLoop library \cite{vanHameren:2010cp}.
Because of very large and complicated expression of the amplitude, we calculate the amplitude numerically before 
squaring it. This requires numerical evaluation of the polarization vectors of gauge bosons.
We choose real basis, instead of helicity basis for the polarization vectors to calculate the amplitude.
This is to reduce the size of compiled program and the time taken in running the code.
We have used RAMBO to generate three particle phase-space for our processes \cite{Kleiss:1985gy}.

  The processes $\gVV$ are leading order one-loop processes. The full amplitude in (1) should be both 
  UV as well as IR finite. However, individual diagrams may be UV and/or IR divergent. IR divergence 
  is relevant to only light quark cases. All these singularities are encoded in various scalar integrals. 
  To make UV and IR finiteness check on our amplitude we have derived all the required scalar integrals 
  (up to box scalar integrals with two massive external legs) analytically following t`Hooft and Veltman 
  \cite{'tHooft:1978xw}. We regulate the UV divergence of the scalar integrals using dimensional 
  regularization and infrared singularities by using small quark mass (the mass regularization).  Our 
  results are in agreement with those given in \cite{Duplancic:2000sk}. We have verified that our amplitude
  is both UV and IR finite and it's gauge invariant with respect to all the gluons. Pentagon and box amplitudes
  are separately gauge invariant with respect to the electro-weak vector bosons. As a consistency check
  we have also verified that the amplitude vanishes in the large quark-mass limit \cite{Appelquist:1974tg}.

\section{Numerical Results}

      We have computed the amplitude numerically using the real polarization
    vector basis. There are 48 polarized amplitudes for the case of $\gamma Z g$ while
    72 for the cases of $ZZg$ and $WWg$. Given the number of diagrams, the number of
    polarization combination and the length of the amplitude, we have to
    run the code in parallel environment using a PVM implementation of the VEGAS
    algorithm \cite{Veseli:1997hr,pvm}. We have used more than 30 cores to run the code 
    in parallel environment. Still it takes more than eight hours to get suitable 
    cross-section for a given process.

       In such loop calculations, there is an issue of numerical instabilities
    due to large cancellations in the evaluation of the scalar and tensor integrals. 
    To minimize this problem, we have used the OneLoop implementation of the
    scalar integrals, $B_{0}, C_{0}, {\rm and}\;  D_{0}$. Even with this implementation, 
    we have faced numerical instabilities in the evaluation of pentagon-tensor integrals
    \footnote{ We would like to emphasize that we have also calculated di-vector 
    boson production via gluon fusion for comparison as discussed below. These processes
    involve tensor integrals up to box type only and we have not faced any numerical instability
    in the calculation. The reduction method adopted in our calculation is indeed very stable numerically.}. 
    This is related to inaccurate evaluation of Gramm determinants for certain phase-space 
    points. These are the phase-space points where linear independency of 
    external momenta (modulo 4-momentum conservation) does not hold to a very good accuracy.
    Indeed they lie on the kinematic boundaries of various channels.
    We judiciously throw away such points by making a gauge invariance test on the full amplitude. 
    Since such points are few, given the nature of the integral evaluation, we don't expect 
    significant impact of this aspect of the calculation. This is a common practice in such
    loop calculations. More complete discussion on this can be found in \cite{Agrawal:2012as,Campanario:2011cs}. 
       
\begin{figure}[ht]
 \begin{minipage}[b]{0.5\linewidth}
\centering
 \includegraphics[width=\textwidth]{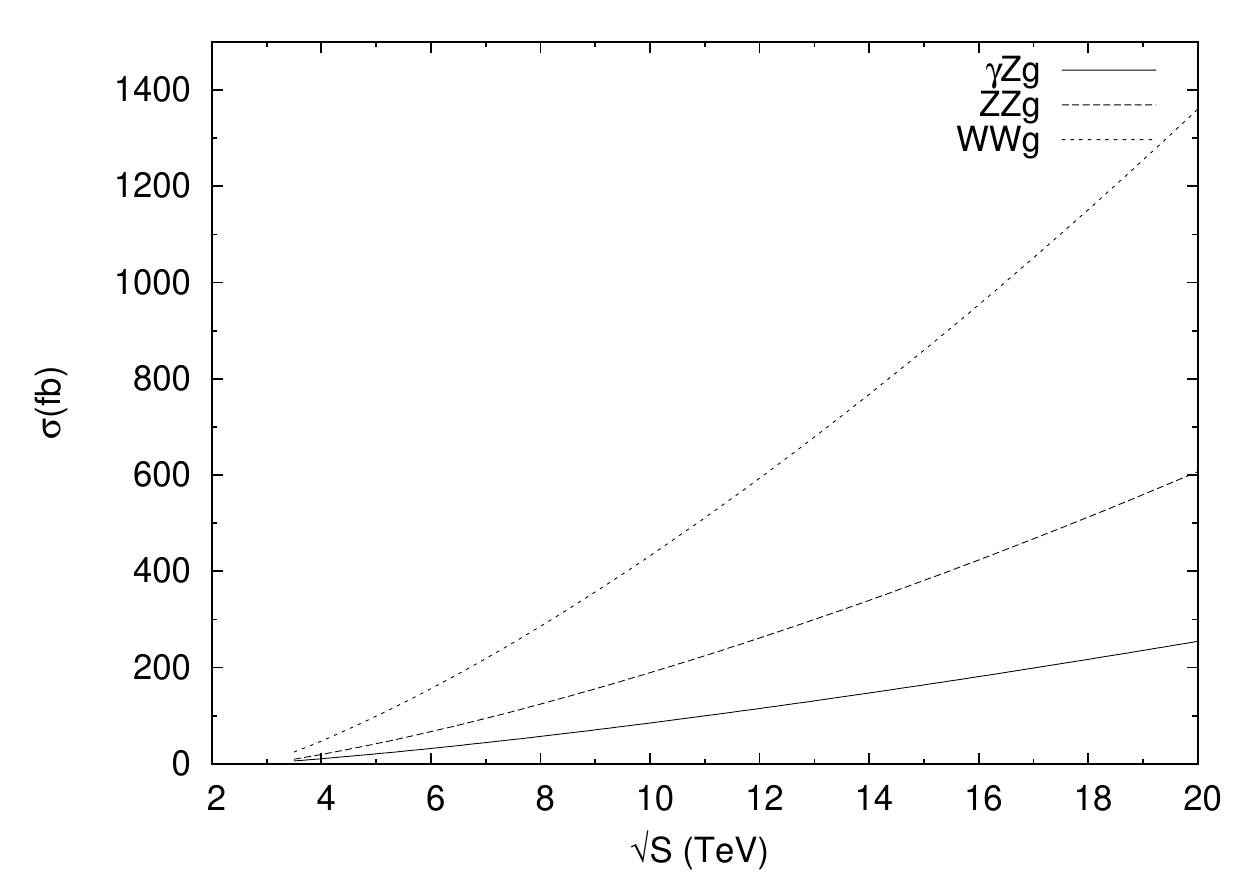}
\caption{Variation of the cross-section with the centre-of-mass energy for $\gVV$.}
\label{fig:sigma_cme}
 \end{minipage}
 \hspace{0.5cm}
 \begin{minipage}[b]{0.5\linewidth}
\centering
\includegraphics[width=\textwidth]{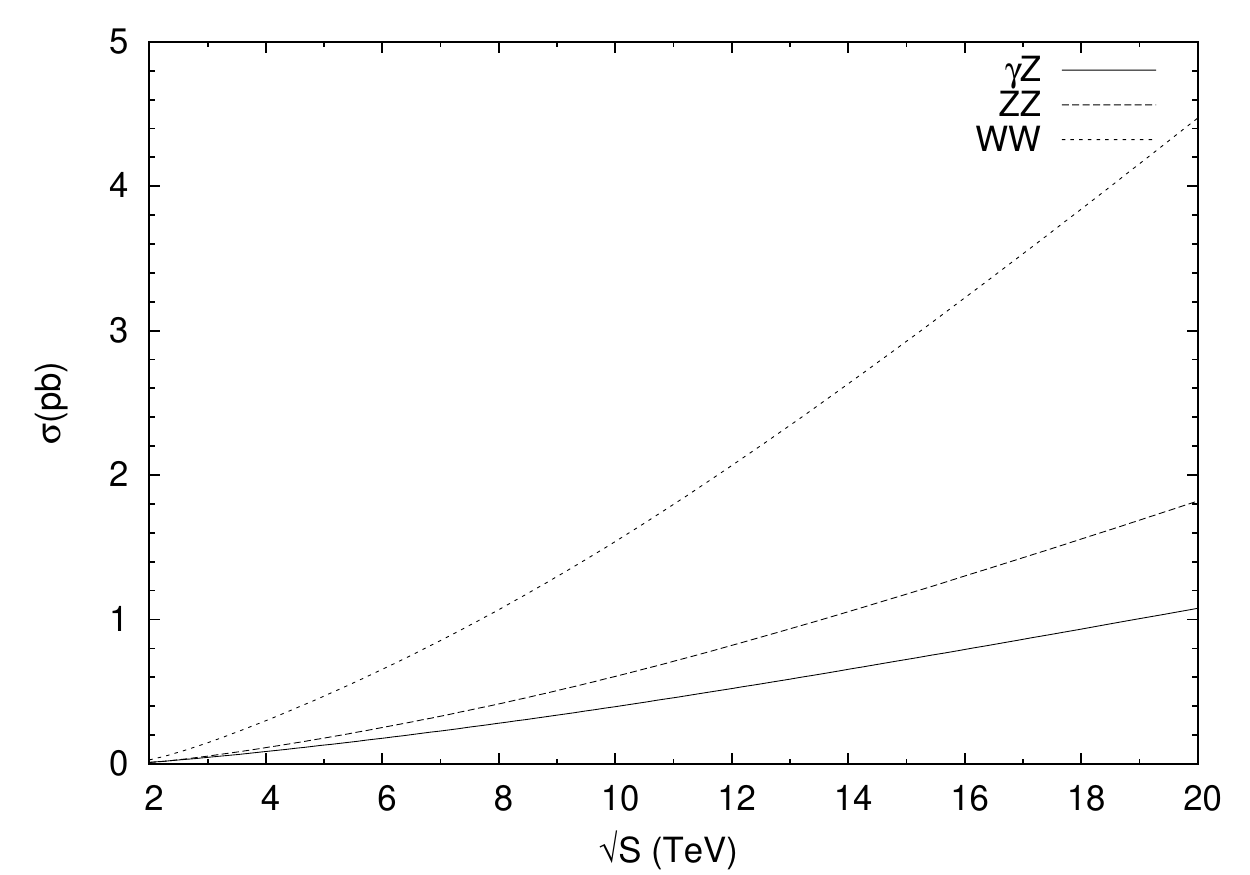}
\caption{Variation of the cross-section with the centre-of-mass energy for $\VV$.}
\label{fig:sigma_cmeVV}
\end{minipage}
\end{figure}

In Fig. (\ref{fig:sigma_cme}), we present the results of the cross-section calculation
for three processes. These results include following kinematic cuts:
$$
P_T^{\gamma,Z,W,j} > 30 \; {\rm{GeV}},\: |\eta^{\gamma,Z,W,j}| < 2.5,\; R(\gamma,j) > 0.6.
$$
We have also chosen factorization and renormalization scales as
$\mu_f = \mu_R = P_T^{\gamma/Z/W}$, as appropriate. We have used CTEQ6M parton distribution 
functions to obtain these results \cite{Nadolsky:2008zw}. We note that at typical LHC
energy, the cross-sections would be of the order of 100 fb. 
For example, at the centre-of-mass energy of the 8 TeV, the cross sections are
46.7 fb, 95.5 fb, and 225.2 fb respectively for the  $\gamma Z g, Z Z g \; {\rm and}\; W^+ W^- g$
production. Therefore, one may expect a few thousand of such events at the end of present run.
But a W/Z boson is observed through its decay channels. If we consider the case when all the
W/Z bosons are seen through their decays to the electron/muon only, then, with $20 \; {\rm fb}^{-1}$
integrated luminosity, the number of events for these processes will be approximately 60, 10, and
220 respectively. However, if we allow one of the W/Z boson to decay hadronically, then the
number of events would increase significantly. At the 14 TeV centre-of-mass energy, the numbers
will be about a factor of three larger.
 The relative behavior of cross-sections of the three processes,
as the center-of-mass energy varies, is quite similar to the case of di-vector boson production 
via gluon fusion as shown in Fig. (\ref{fig:sigma_cmeVV}). This common behavior is mainly due to 
the relative couplings of the processes listed above in (3). We find that at 14 TeV the cross-sections
for our processes are $20-30 \%$ of those for corresponding di-vector boson production 
(without jet) processes. We can also compare the contribution of the loop processes
with the tree-level processes. We find that the processes $g g \to \gamma Z g,
W^{+}W^{-}g$ make about $4-5 \%$ contribution to the processes $ p p \to \gamma Z j,
W^{+}W^{-} j$, while $g g \to ZZ g$ makes a contribution of about $10-15 \%$ to the
$p p \to ZZ j$ process. Here `$j$' stands for a jet. This is quite similar to the case 
of di-vector boson production. Tree level estimates are obtained using MadGraph \cite{Alwall:2011uj}.
We have also compared our results for the $W^{+}W^{-} j$ production
with those of {\it Melia et al.} \cite{Melia:2012zg}. Though they have considered the leptonic 
decays of $W$-bosons and kinematic cuts, choice of scales and parton distributions etc are quite different, the 
percentage contribution of gluon-gluon channel as compared to the LO cross-section is same within the 
allowed range of uncertainty, {\it i. e.} $4-5 \%$. The values of the cross-section are more strongly
dependent on the values of parameters and kinematic-cuts; still two results are similar
if we take into account quoted uncertainties and branching ratios.
The contribution of these gluon fusion processes can be even larger in appropriate kinematic regime.

\begin{figure}[ht]
 \begin{minipage}[b]{0.5\linewidth}
\centering
 \includegraphics[width=\textwidth]{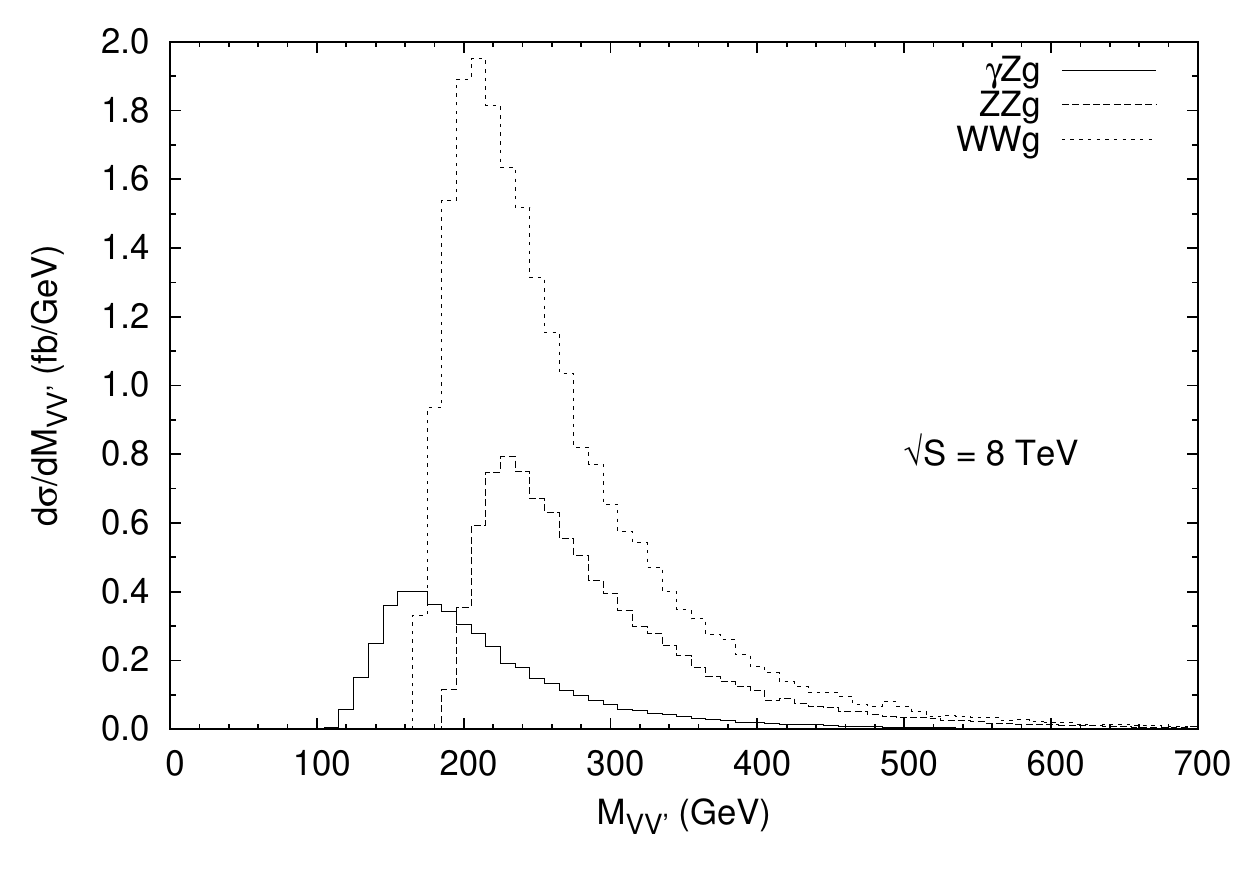}
\caption{Invariant Mass distribution of vector boson pairs at 8 TeV.}
\label{fig:m12-8}
 \end{minipage}
 \hspace{0.5cm}
 \begin{minipage}[b]{0.5\linewidth}
\centering
 \includegraphics[width=\textwidth]{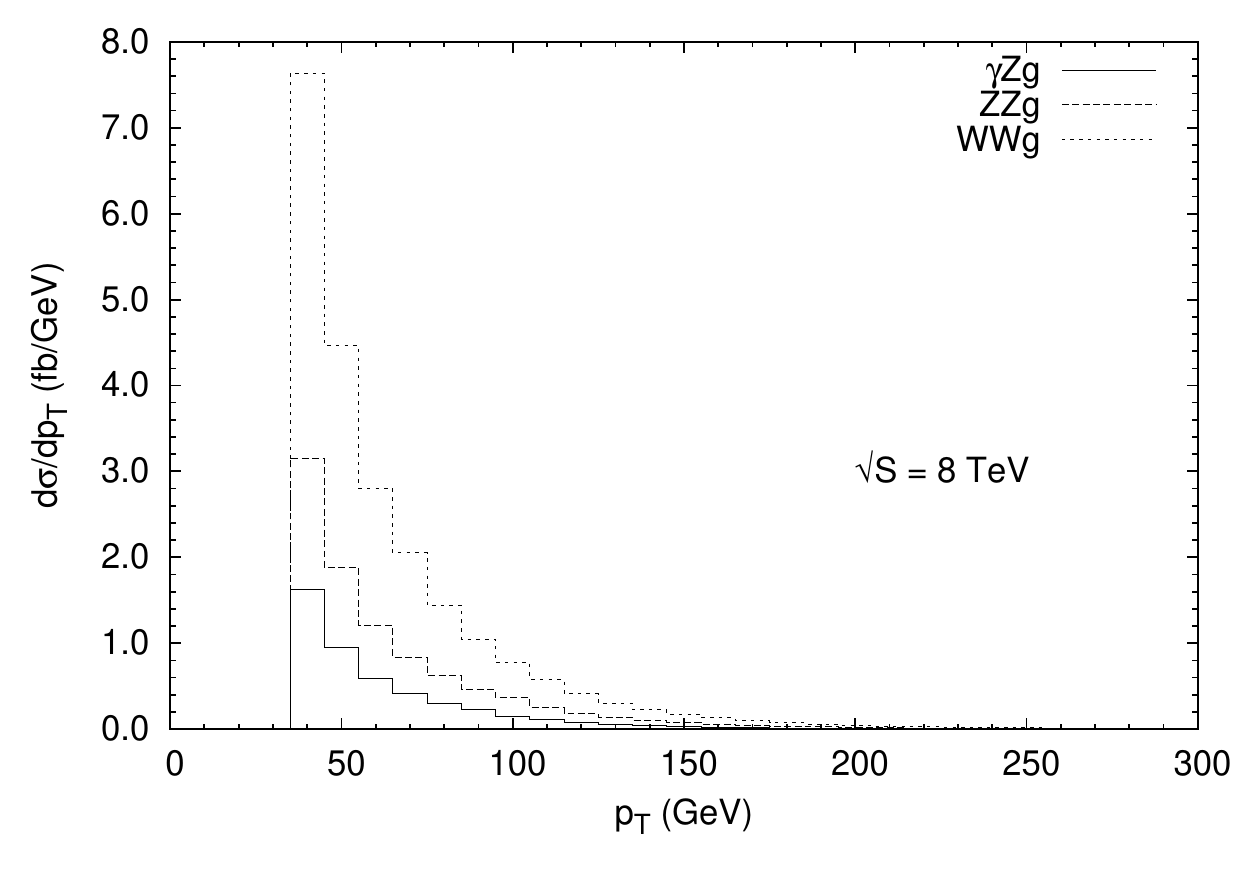}
\caption{Transverse momentum distribution of Gluons at 8 TeV.}
\label{fig:ptg-8}
 \end{minipage}
\end{figure}

\begin{figure}[h!]
\begin{center}
 \includegraphics[width=0.6\textwidth]{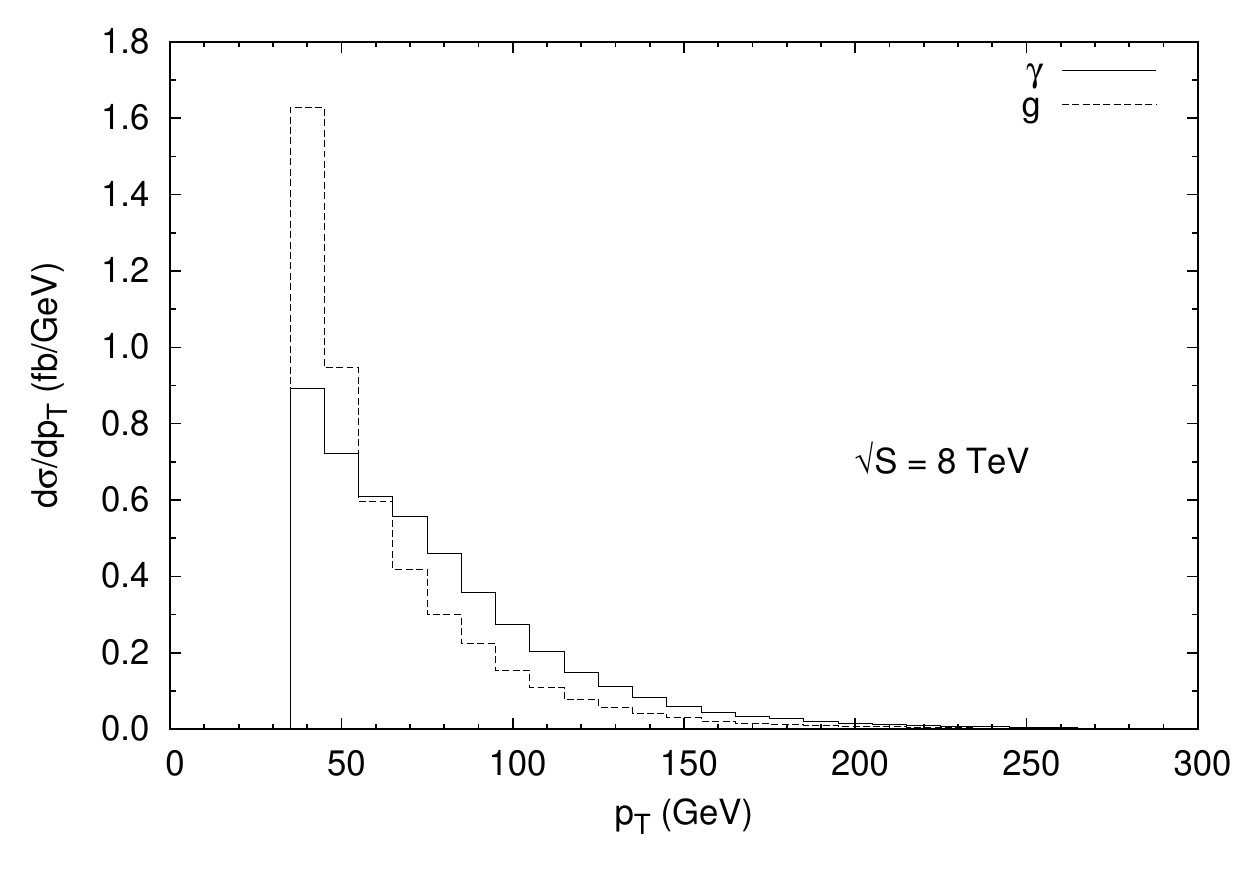}
\caption{Comparison of $P_T$ distributions of $\gamma$ and $g$ for $\gamma Z g$ production at 8 TeV.}
\label{fig:pt-pg-8}
\end{center}
\end{figure}

We now discuss few kinematic distributions at 8 TeV collider centre-of-mass energy. 
These distributions remain same, characteristically, at 14 TeV centre-of-mass energy.
The invariant mass distributions for pairs of vector boson are shown in Fig. (\ref{fig:m12-8}).
The positions of the peaks are related to the masses of the vector bosons in each case. In Fig. (\ref{fig:ptg-8})
the transverse momentum distributions for gluon is given for three processes.
The major contribution to the cross-section comes from 
low $P_T$ region, as we expect. The cross-section is very sensitive to $P_T$ cut on gluon
as it can come from one of the incoming gluons in the box contribution.
We have compared the $P_T$ distributions of $\gamma$ and $g$ for the case of $\gamma Z g$ in Fig. (\ref{fig:pt-pg-8}).
The rapid fall of $P_T^g$ distribution compared to that of $P_T^\gamma$ can be seen clearly.
This is because the photon is coming from the quark loop,
rather than from a quark-leg (as in the case of tree-level process). A larger cut on this transverse
momentum will enhance the relative contribution of the loop process.

\section{Conclusions}

    We have presented the results of the cross-section calculations 
    for the processes $g g \to \gamma Z g, Z Z g \;{\rm and} \; W^{+} W^{-} g$.
    These are one-loop processes and proceed 
    via pentagon and box diagrams. We have kept only $\gamma Zg$ like contributions
    for the cases of $ZZg$ and $WWg$. There is decoupling of the top-quark.
    For the $WWg$ case,
 the contribution from the first two generations of the quark is included.
     We have made a number of checks 
    on our calculation. We have verified the cancellation of UV and mass 
    singularities. We have also checked gauge invariance with respect to
    all the gauge particles. Axial-vector contribution does not 
    interfere with the vector contribution and it is separately gauge invariant.
    We have observed qualitative similarity of these processes with the 
    corresponding di-vector boson production cases.
    All these processes have already generated several hundred events at the LHC. 
    In appropriate kinematic regime one should be able to observe these processes. 
    The contribution of the process $g g \to Z Z g$ is specially significant. 
    It may have bearing on the searches of the Higgs boson and new physics scenarios 
    that lead to two Z-boson in the final state.
    Detailed and more complete study of these processes will be reported in future 
    publications. \\

AS would like to acknowledge fruitful discussions 
with R. Frederix and D. Zeppenfeld. He would also
like to thank Dr. S. Veseli for his help with AMCI package.

\end{document}